\documentclass[submission,copyright,creativecommons]{eptcs}
\providecommand{\event}{}
\usepackage{breakurl}  
\usepackage{graphicx}
\usepackage{graphics}
\usepackage{subfigure}
\usepackage{epsfig}
\title{Components Interoperability through \\Mediating Connector Patterns\thanks{This work is an extension of our previous work \cite{SI_ECSA_10}.}  \thanks{
This work is partly supported by the {\sc Connect} European Project No 231167.
}}

\author{Romina Spalazzese \qquad\qquad Paola Inverardi
\institute{Universit\`a degli Studi dell'Aquila\\
via Vetoio I-67100 L'Aquila, Italy}
\email{\quad romina.spalazzese@di.univaq.it \quad\qquad paola.inverardi@di.univaq.it}
}
\def\titlerunning{Components Interoperability through Mediating Connector Patterns}
\def\authorrunning{Romina Spalazzese and Paola Inverardi}
\begin{document}
\maketitle

\begin{abstract}
A key objective for ubiquitous environments is to enable system interoperability between system's components that are highly heterogeneous. In particular, the challenge is to embed in the system architecture the necessary support to cope with behavioral diversity in order to allow components to coordinate and communicate. The continuously evolving environment further asks for an automated and on-the-fly approach. In this paper we present the design building blocks for the dynamic and on-the-fly interoperability between heterogeneous components.
Specifically, we describe an Architectural Pattern called \textit{Mediating Connector}, that is the key enabler for communication. In addition, we present a set of \textit{Basic Mediator} Patterns, that describe the basic mismatches which can occur when components try to interact, and their corresponding solutions.
\end{abstract}

\section{Introduction}
\label{SEC:intro}
A multitude of heterogeneous networked devices are today embedded in the Ubiquitous networked environment \cite{WEISER} where a key objective is to enable system interoperability between system's components that are highly heterogeneous.

Tremendous work has been done in the middleware field to enable diverse networked systems to actually work together, especially to respond to the new needs introduced by the ubiquitous environments.
Devices need to automatically detect services available in the environment and adapt their own communication protocols to interoperate with them since networked applications are developed on top of diverse middleware.
However, the proposed solutions only address middleware-layer interoperability letting still open the interoperability at application-layer that calls for \textit{mediating connectors} or \textit{mediators} for short.

The mediator concept was initially introduced to cope with the integration of heterogeneous data sources \cite{WIEDERHOLD,WIEDERHOLD2}, and as design pattern \cite{GOF}. Moreover, mediators have received an increasing attention within the Web Services and Semantic Web contexts. With the significant development of Web technologies, a big effort has been devoted to mediate many heterogeneity dimensions, spanning~\cite{STOLLBERG}: terminology, representation format and transfer protocols, functionality and application-layer protocols.
Protocol mediation is further concerned with behavioral mismatches that may occur during interactions. Other approaches that share the same formal settings as protocol mediation have been proposed to solve mismatches in the field of supervisory control theory of discrete event systems \cite{KUMAR}. More recently, in the field of software architectures ad hoc wrappers have been proposed to address communication problems as for example data translation, connectors combination, role addition, extra functional deficiencies \cite{SPITZ}. A lot of work has been also devoted to behavioral adaptation, \cite{BENATALLAH2,DUMAS,TSE_CANAL} to mention few.
Automated mediation has deserved attention, especially in the context of Web services and Semantic Web technologies \cite{VACULIN07,VACULIN08,SEMI07,WILLIAMS}.
More recently the challenge is to provide general solutions to the behavioral diversities at runtime and on-the-fly, to respond to the continuous evolution of the environment \cite{CONNECT}.

All these problems call for an approach to protocol mediation based on the categorization of the types of protocol mismatches that may occur and that must be solved in order to provide corresponding solutions to these recurring problems.
This immediately reminds of patterns \cite{ALEXANDER,BUSCHMANN,PARIS,GOF}. In this paper we present a set of design building blocks for the interoperability between heterogeneous components.
A catalogue of problems and their related solutions would not solve all the possible mismatches but it would certainly facilitate the solution.

The contributions of this paper are:
(1) an Architectural Pattern called \textit{Mediating Connector}, that is the key enabler for communication; (2) a set of \textit{Basic Mediator} Patterns that describe: (i) the basic mismatches which can occur while components try to interact, and (ii) their corresponding solutions. This work extends a preliminary version \cite{SI_ECSA_10} with: the introduction of a related work section, the introduction of a detailed description of basic mismatch patterns and a extended description of the example including a new section on the patterns application on it. This work is part of the {\sc Connect}  European Project \cite{CONNECT} where we investigate an overall framework for the seamless networking of heterogeneous systems.

The remaining of the paper is organized as follows. In Section \ref{SEC:example}, we introduce a motivating example for protocol mediation. In Section \ref{SEC:approach}, we describe a pattern-based approach which we envision for the automatic synthesis of mediating connectors for the ubiquitous networked environment. We illustrate the Mediating Connector Architectural Pattern in Section \ref{SEC:arch_pattern}. In Section \ref{SEC:basic_patterns}, we illustrate the Basic Mediator Patterns including the basic interoperability mismatches, which can occur when two heterogeneous components try to interoperate, and their respective solutions.
In Section \ref{SEC:application}, we show the application of the mediator patterns to the example to solve the interoperability mismatches. Then, we discuss related work in Section \ref{SEC:related} and we conclude, in Section \ref{SEC:conclusion}, by also outlining future work.

\section{Example Motivating the Mediation need}
\label{SEC:example}
%
To better illustrate \textit{protocol mediation}, and to make the
underlying problem more concrete, in the following we describe the example used in \cite{SII_WICSA_ECSA09} where we have been studying the problem and first results on the theory our approach is based on appeared. An extended and more complete version of the theory can be found \cite{ISOLA_10_THEO}.
We consider the simple yet challenging example of instant messaging. Various instant messaging systems are now in use, facilitating communications among people. However, although those systems implement similar functionalities, end-users need to use the very same system to communicate due to behavioral mismatches of the respective protocols.

\begin{figure}
\centering
\subfigure[Windows Messenger protocol]{
        \epsfig{file=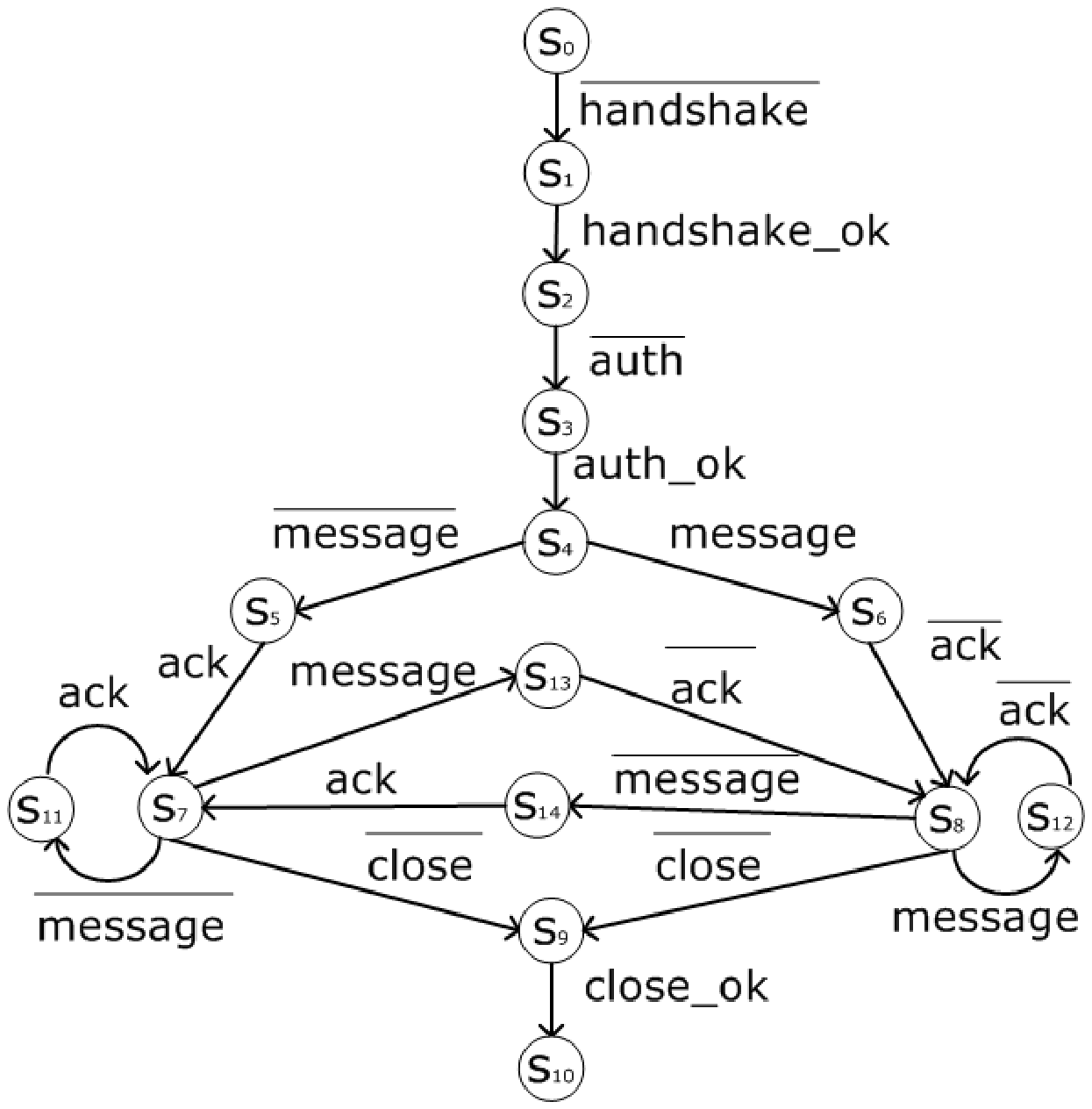, width=2.0in}
        \label{FIG:msn_2}}
    \subfigure[Jabber protocol]{
        \epsfig{file=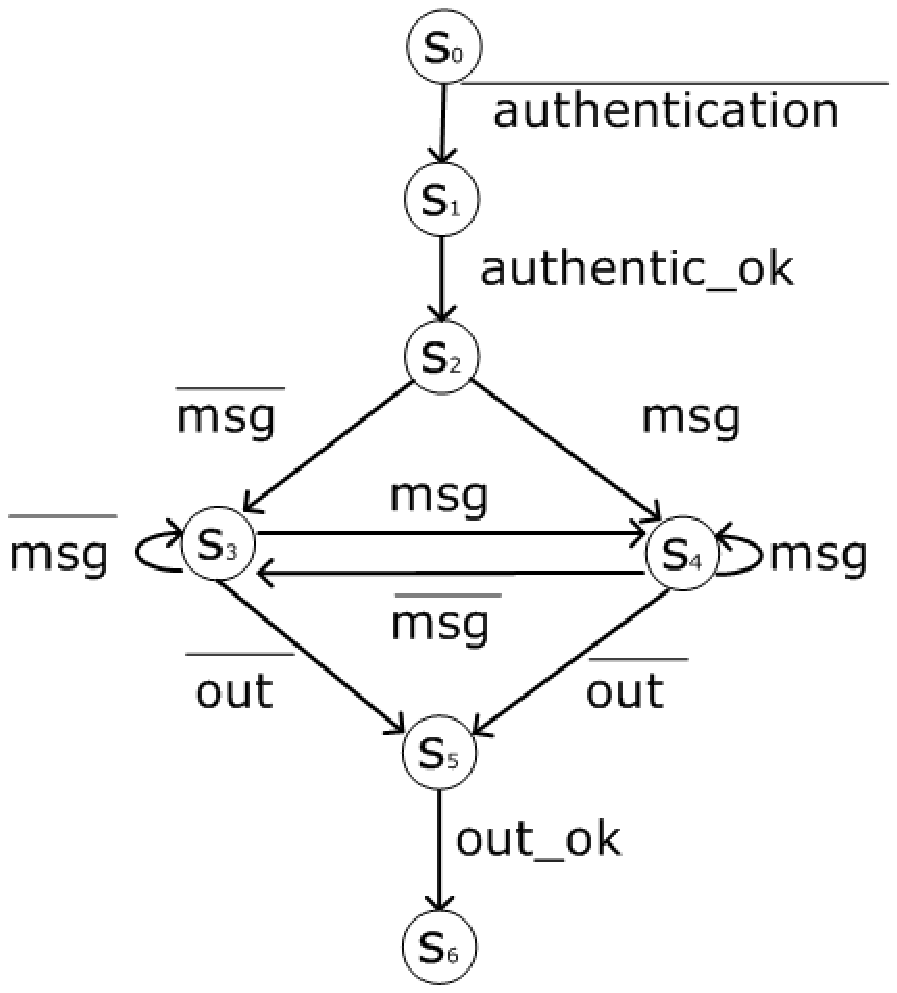, width=1.6in}
        \label{FIG:jabber}}
\caption{Behavioral modeling of two instant messaging protocols}
\label{FIG:twoIM}
\end{figure}

For instance, consider Windows Messenger\footnote{Windows Live Messenger, http://www.messenger.it/}(WM), and Jabber Messenger\footnote{Jabber Software Foundations, http://www.jabber.org/}(JM).
Figure \ref{FIG:twoIM} models the respective behaviors of the
associated protocols using Labeled Transition Systems (LTS)
\cite{LTS}. LTSs constitute a widely used model for concurrent computation and are often used as a semantic model for
formal behavioral languages such as process algebras. We use the
usual convention that actions with overbar denote output actions
while the ones with no overbar denote input actions. These systems
should be able to interoperate since they both amount to supporting authentication with their servers and then message
exchanges among peers. However mediating their respective protocols
to achieve interoperability is far from trivial, especially if one
wants to achieve a general solution.

An effort has been done in \cite{CROSSTALK} to mediate instant
messaging protocol mismatches allowing communication between any two
clients. Unfortunately, the proposed solution requires the
implementation of the translation from any client protocol (to be
supported) to a reference exchange protocol to be given, and vice
versa. This obviously affects the generality and the automation of
the approach.
Several works have devoted attention to protocol mediation in the
context of Web services using patterns \cite{CIMPIAN},\cite{BENATALLAH},\cite{CIN2}. In particular the
latter provides tools to developers to help them to identify
protocol mismatches and to compose mediators. However, this remains
quite limited with respect to enabling interoperability in today's
networking environments that are highly dynamic. Indeed, mediators
need to be synthesized on demand so as to allow interactions with
networked systems that are not known in advance. Such a concern is
in particular recognized by the Web service research community that
has been studying solutions to the automatic mediation of business
processes in the recent years.

\section{A Pattern-based Approach for Interoperability Mismatches}
\label{SEC:approach}
In this section we describe our proposal for an automated pattern based approach. It is based on first results describing a theory of mediators at a high level \cite{SII_WICSA_ECSA09}. For the sake of this paper, we make some assumptions and we investigate the related underling research problems as part of {\sc Connect}. We assume to know the interaction protocols run by two networked components as Labeled Transition Systems (LTS) and the components' interfaces with which to interact as advertised or as result of learning techniques \cite{CONNECT_PAPER,STRAWBERRY}.
We also assume a semantic correspondence between the messages exchanged among components exploiting ontologies.
The first step is to establish whether the components are \textit{potentially compatible}, that is, if it makes sense for them to \textit{interoperate} through the Mediating Connector. This amounts to understand if the components share \textit{intents} (traces), i.e., if, modulo some adaptation, they show complementary sequences of messages visible at interface level. To do this we envision (1) a decomposition strategy/tool to decompose the whole components' behavior (LTS) into {\it elementary behaviors} (traces) representing {\it elementary intents} of the components and (2) an automatic analyzer to identify mismatches between elementary behaviors of the different components as done in other research areas \cite{BENATALLAH2,NEJATI_ICSE07}. Once discovered the components compatibility, solving their interoperability means solving the behavioral mismatches that they exhibit. Then it is necessary to: (3) define a mismatches manager to solve the identified mismatches between elementary behaviors; (4) define a composition approach to build elementary mediating behaviors (mediating traces) based on the identified mismatches and their relative solutions; (5) define a composition strategy to build a mediating connector's behavior starting from the elementary mediating behaviors.

The above described approach is far from trivial, especially to achieve automatically. However, in the following we show its feasibility. To address steps (1) and (5) the approach makes use of a compositional strategy to decompose components interaction protocols into traces and compose mediating connectors interaction protocol from mediating traces respectively.
Furthermore, we describe six Basic Mediator Patterns that are the building blocks on which the steps (2), (3), and (4) can be built upon.

\section{Mediating Connector Architectural Pattern}
\label{SEC:arch_pattern}
The interoperability problem between diverse components populating the ubiquitous environment and its related solution is characterized as a \textit{Mediating Connector Architectural Pattern} basing on the template used in \cite{BUSCHMANN} that contains the following fields: Name, Also Known As, Example, Context, Problem, Solution, Structure, Dynamics, Implementation, Example Resolved, Variants, Consequences. The Mediating Connector is a behavioral pattern and represents the architectural building block embedding the necessary support to dynamically cope with components' behavioral diversity.
\\
\noindent\textbf{Name. Mediating Connector.}
\\
\noindent\textbf{Also Known As.} Mediator.
\\
\noindent\textbf{Example.} Consider the ubiquitous environment that
embeds networked devices from a multitude of applications domain,
for example consumer electronics or mobile and personal computing
devices. Suppose that \textit{potentially compatible} applications
running on various devices want to \textit{interoperate}.
Potentially compatible applications are applications that may share
some \textit{intent} resulting in complementary portions of their
\textit{interaction protocols}, i.e.,  complementary sequences of
messages visible at interface level. In principle those applications
should be able to interoperate, but because of some
\textit{behavioral differences} that they exhibit, they are not
compatible. With interoperate we mean \textit{coordinate} and
\textit{communicate (i.e. \textit{synchronize})}. For example,
consider the instant messaging applications described in Section
\ref{SEC:example}. Users of different messengers may want to
communicate and in principle this should be possible since the two
different messengers implement similar functionalities. However they
do it in different ways and this prevents the communication.
\\
\noindent\textbf{Context.} The environment is distributed and
changes continuously. Heterogeneous (mismatching) systems populating the
environment require seamless coordination and communication.
\\
\noindent\textbf{Problem.} In order to support existing and future
systems' interoperability, some means of mediation is required.
%
From the components' perspective, there should be no difference
whether interacting with a peer component, i.e, using the very same
interaction protocol, or interacting through a mediator with another
component that uses a different interaction protocol. The component
should not need to know anything about the protocol of the other one
while continuing to "speak" its own protocol.

Using the Mediating Connector, the following \textit{forces}
(aspects of the problem that should be considered when solving it \cite{BUSCHMANN}) need to be balanced: (a) the different components should continue to use their own interaction protocols. That is components should interact as if the Mediating Connector were transparent; (b) the following \textit{basic interaction protocol mismatches} should be solved in order for a set of components to coordinate and communicate
(a detailed description of these mismatches is given within Section \ref{SEC:basic_patterns}): 1) Extra Send/Missing Receive Mismatch; 2) Missing Send/Extra Receive Mismatch; 3) Signature Mismatch; 4) Ordering Mismatch; 5) One Send-Many Receive/Many Receive-One Send Mismatch; 6) Many Send-One Receive/One Receive-Many Send Mismatch

\noindent\textbf{Solution.} The introduction of a Mediating
Connector to manage the interaction behavioral differences between
potentially compatible components. The idea behind this pattern is
that, by using the Mediating Connector, components that would need
some interaction protocol's adaptation to become compatible, and
hence to interoperate, are able to coordinate and communicate
achieving their goals/intents without undergoing any modification.

\begin{figure}[htbp]
\centering \epsfig{file=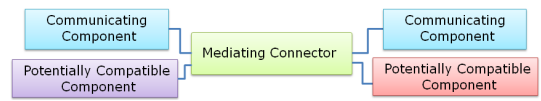,
width=2.8in}
 \caption{Entities involved in a mediated system}
\label{FIG:structure}
\end{figure}

The Mediating Connector is one (or a set of) component(s) that
manage the behavioral mismatches listed above. It directly
communicates with each component by using the component's proper protocol.
The mediator forwards the interaction messages from one component to
the other by making opportune translation/adaptation of protocols
when/if needed.
\\
\noindent\textbf{Structure.}
The Mediating Connector Pattern comprises three types of participating components: communicating components, potentially compatible components and mediators.
\\
\noindent The {\em communicating components} implement already compatible components, i.e. components able to interact and evolve following their
usual interaction behavior. The {\em potentially compatible components} implement the application level entities (whose behavior, interfaces' description
and semantic correspondences are known). Each component wants to
reach its intents by interacting with other components able
to satisfy its needs, i.e. required/provided functionalities. However the components are unable to directly interact because of protocol mismatches. Thus,
the potentially compatible components can only evolve following their
usual interaction behavior, without any change. The {\em mediators} are entities responsible for the mediated
communication between the components. This means that the role of
the mediator is to make compatible components that are mismatching. That is, a mediator must receive and
properly forward requests and responses between potentially
compatible components that want to interoperate.
Figure \ref{FIG:structure} shows
the object involved in a mediated system.
%

\begin{figure}[htbp]
\centering \epsfig{file=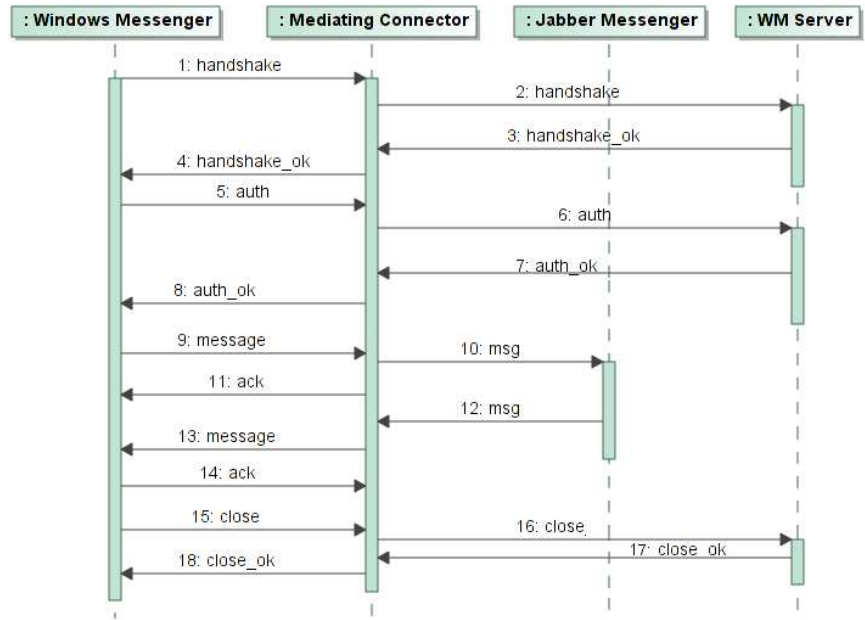,
width=4.5in}
 \caption{Scenario on the relevant operation of a Mediator}
\label{FIG:sequence}
\end{figure}

\noindent\textbf{Dynamics.}
Figure \ref{FIG:sequence} illustrates
the interactions between three components and
one mediator belonging to the messengers example.
Triggered by a user, the WM protocol (Figure \ref{FIG:msn_2}) performs one of its possible behavior: it authenticates after an handshake, sends/receives several messages, and closes.
The mediator should: (1) forward the handshake and authentication messages as they are between the WM and its authentication server (communicating components), (2) translate and forward messages between the WM and JM (potentially compatible components), and (3) forward the closing messages as they are between the WM and its server (communicating components). With the term ``translation'' we mean not just a language translation but also a ``behavioral translation'' (see Section \ref{SEC:basic_patterns} for details).

%
\noindent\textbf{Implementation.}
The implementation of this pattern implies the definition of an approach/tool (we have proposed one in Section \ref{SEC:approach}) to automatically synthesize the behavior of the Mediating Connector which allows the potentially compatible components to interoperate mediating their interactions.
\\
\noindent\textbf{Example Resolved.}
The Mediating Connector's concrete protocol for the example is shown in Figure \ref{FIG:connector}.
Once established that they are potentially compatible (i.e. they have some complementary portion of interaction protocols), the mediating connector manages
the components' behavioral mismatches allowing them to have a mediated coordination and communication.
\\
\noindent\textbf{Variants.} Distributed Mediating Connector. It is possible to implement this pattern either as a centralized component or as distributed components, that is by a number of smaller components. This introduces a synchronization issue that has to be taken into consideration while building the mediator behavior.
\\
\noindent\textbf{Consequences.} The main \textit{benefit} of the Mediating Connector Pattern is that it allows interoperability between components that otherwise would not be able to do it because of their behavioral differences. These components do not use the very same observable protocols and this prevents their cooperation while, implementing similar functionalities, they should be able to interact.
The main \textit{liability} that the Mediating Connector Pattern imposes is that systems using it are slower than the ones able to directly interact because of the indirection layer that the Mediating Connector Pattern introduces. However the severity of this drawback is mitigate and made acceptable by the fact that such systems, without mediator, are not able at all to interoperate.
\\

\section{Basic Mediator Patterns}
\label{SEC:basic_patterns}
In the previous sections we characterized the Mediating Connector pattern and we sketched an approach for the automatic synthesis of its behavior.

In this section, we concentrate on six finer grain \textit{Basic Mediator Patterns} which represent a systematic approach to solve interoperability mismatches that can occur during components' interaction. The Basic Mediator Patterns are constituted by basic interoperability mismatches with their corresponding solutions and are: (1) Message Consumer Pattern, (2) Message Producer Pattern, (3) Message Translator Pattern, (4) Messages Ordering Pattern, (5) Message Splitting Pattern, (6) Messages Merger Pattern.

The mismatches, inspired by service composition mismatches, represent send/receive problems that can occur while synchronizing two traces. In this paper we are not considering parameters mismatches which are extensively addressed elsewhere \cite{DATA}.
%
\begin{figure}[htbp]
\centering \epsfig{file=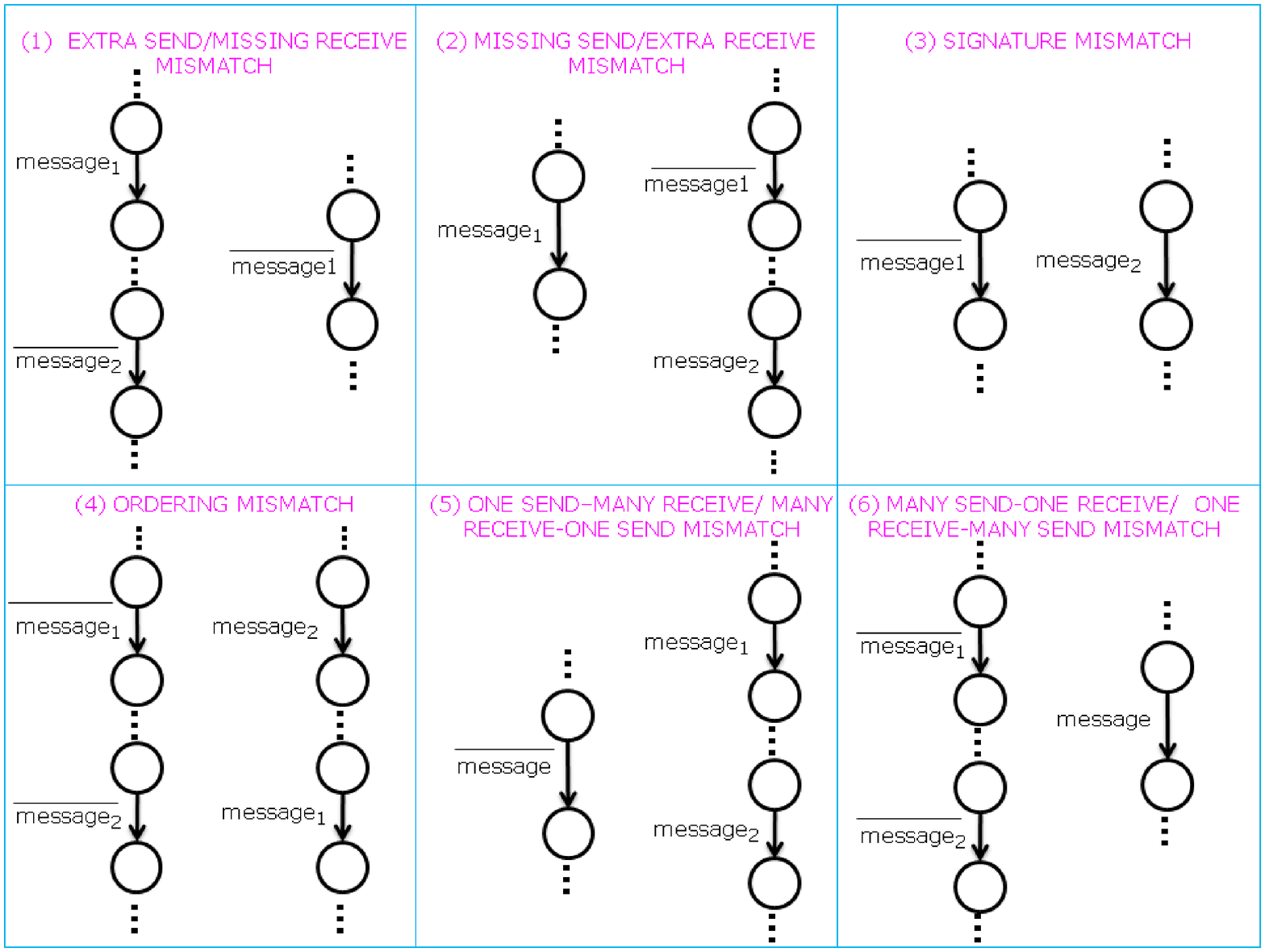,
width=4.5in}
 \caption{Basic interoperability mismatches}
\label{FIG:basic_mismatches}
\end{figure}
Figure \ref{FIG:basic_mismatches} shows the basic interoperability mismatches that we explain in detail in the following.
For each basic interoperability mismatch, we consider two traces (left and right) coming from two potentially compatible components.
All the considered traces are the most elementary with respect to the messages exchanged and only visible messages are shown.

\begin{figure}[htbp]
\centering \epsfig{file=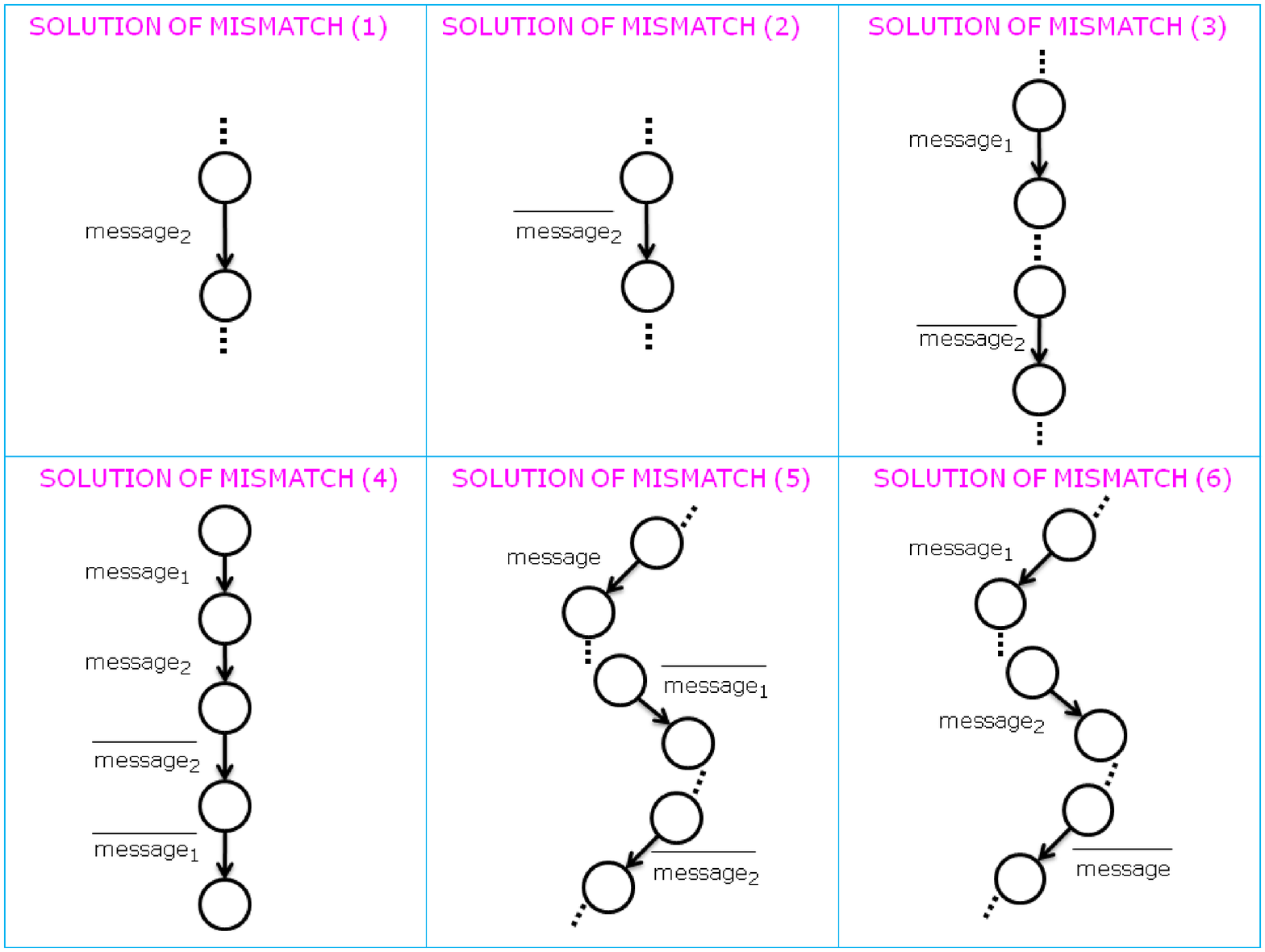,
width=4.5in}
 \caption{Basic solutions for the basic mismatches}
\label{FIG:basic_mediator_patterns}
\end{figure}

It is obvious that, in real cases, the traces may also contain portions of behavior already compatible (abstracted by dots in the figure) and may amount to any combination of the presented mismatches. Then an appropriate strategy to detect and manage this is needed.
The considered basic mismatches are addressed by the basic solutions (elementary mediating behaviors) illustrated in Figure \ref{FIG:basic_mediator_patterns} where only their visible messages are shown (messages that they exchange with the components).

%

The six Basic Mediator Patterns share the context, i.e., the situation in which they may apply and have a unique intent.
\\
\noindent\textbf{Context.} Considering two traces (left and right) expressing similar complementary functionalities. Focus on one of their semantically equivalent elementary actions.

\begin{figure}[htbp]
\centering \epsfig{file=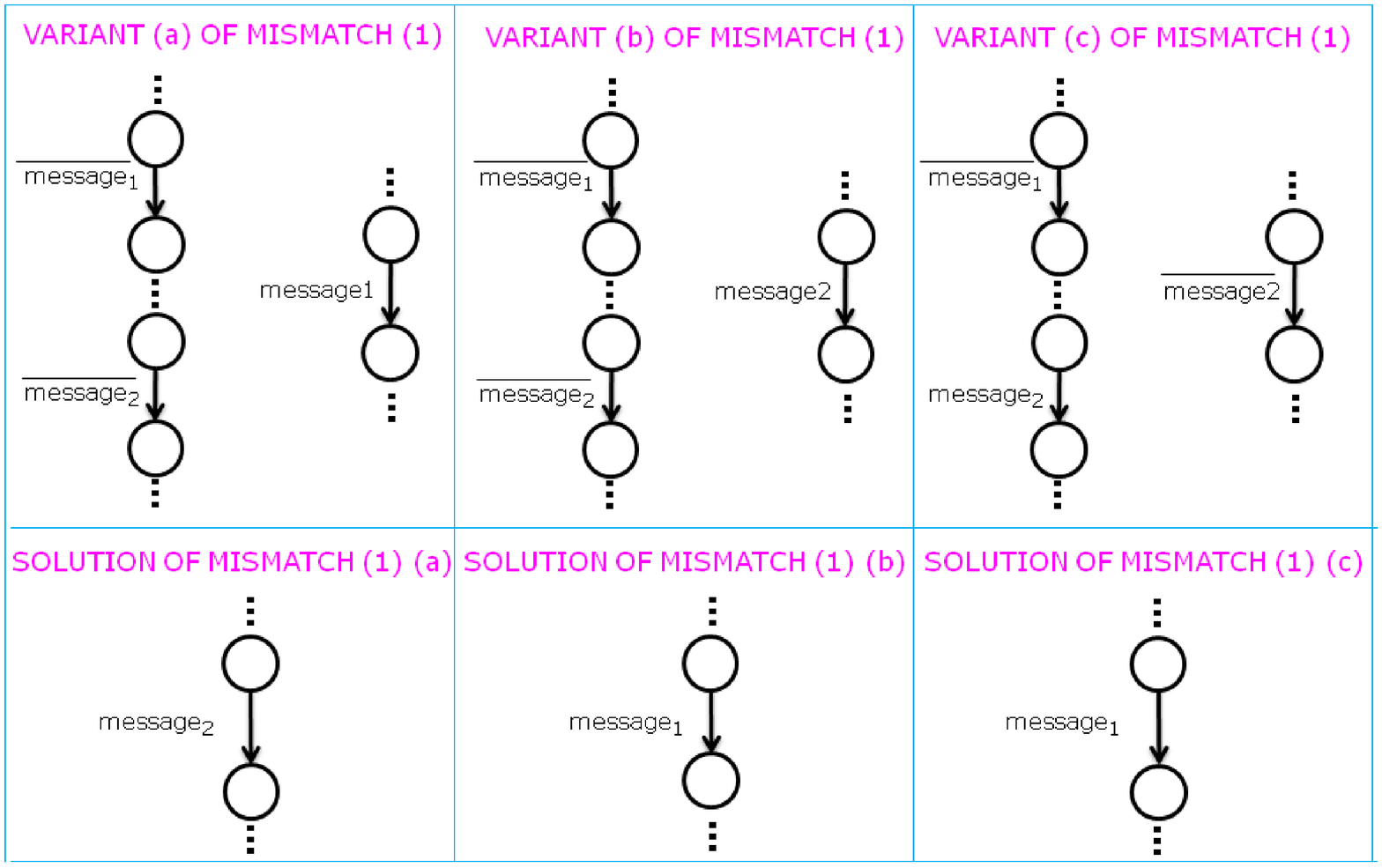,
width=4.0in}
 \caption{Variants of the Basic Mediator Pattern (1)}
\label{FIG:var_basic_mismatch_1}
\end{figure}

\begin{figure}[htbp]
\centering \epsfig{file=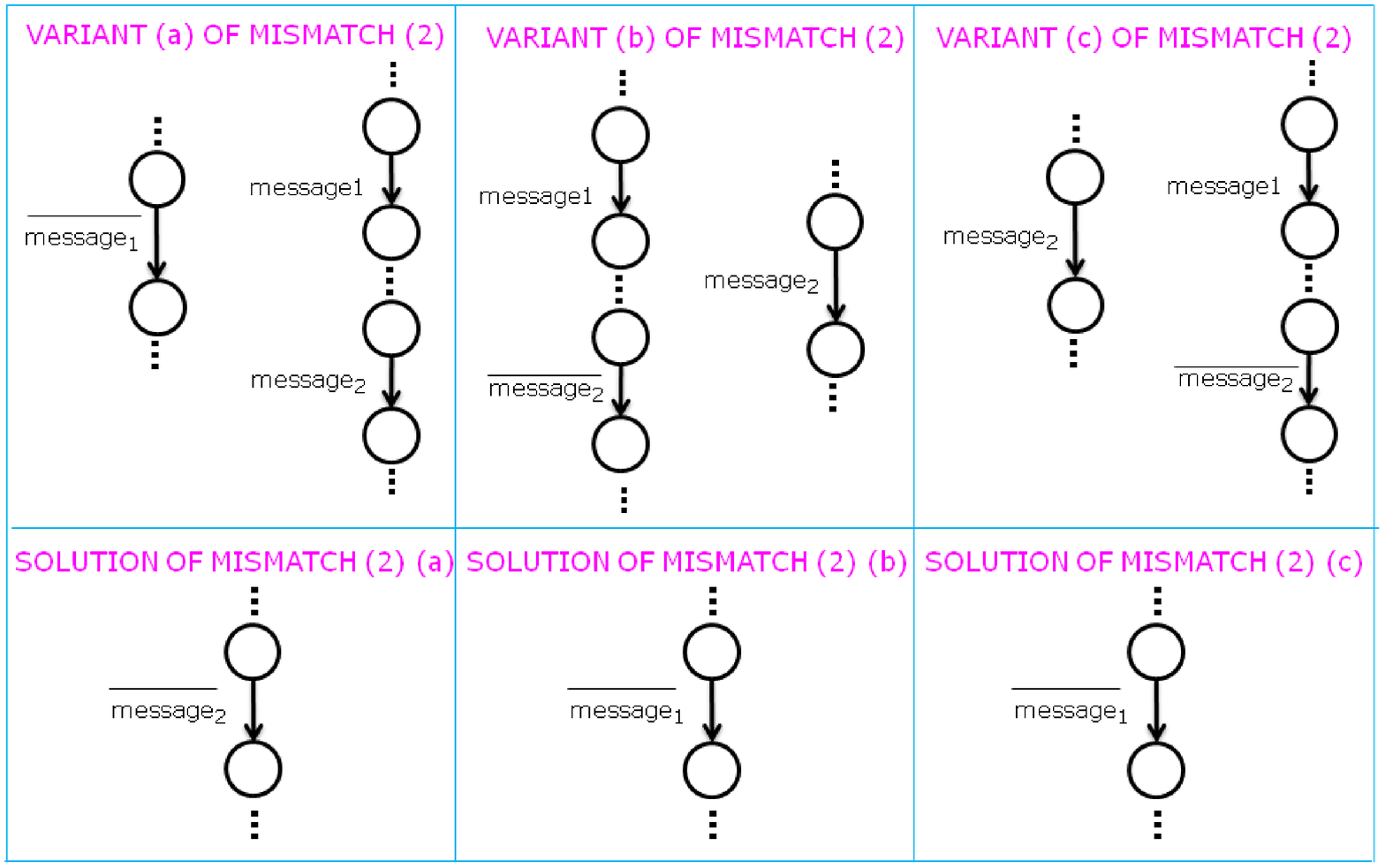,
width=4.0in}
 \caption{Variants of the Basic Mediator Pattern (2)}
\label{FIG:var_basic_mismatch_2}
\end{figure}

\begin{figure}[htbp]
\centering \epsfig{file=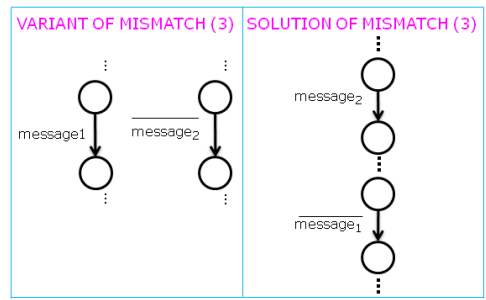,
width=2.5in}
 \caption{Variants of the Basic Mediator Pattern (3)}
\label{FIG:var_basic_mismatch_3}
\end{figure}

\begin{figure}[htbp]
\centering \epsfig{file=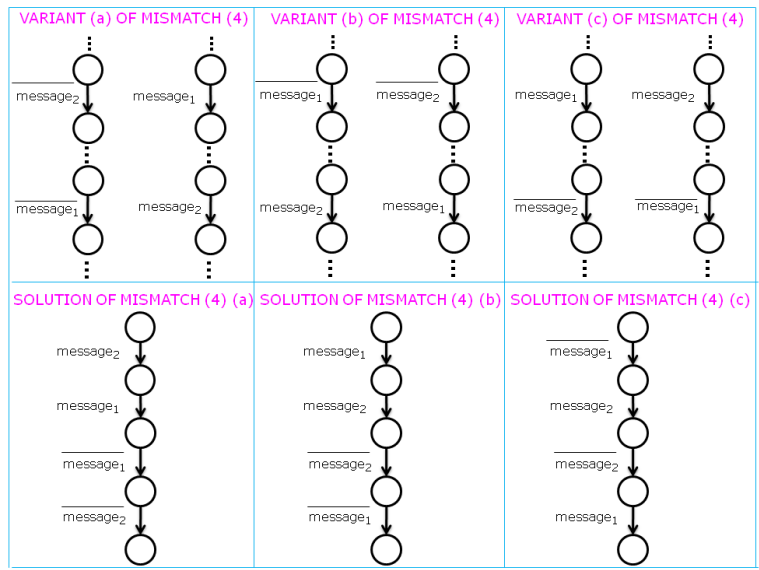,
width=4.0in}
 \caption{Variants of the Basic Mediator Pattern (4)}
\label{FIG:var_basic_mismatch_4}
\end{figure}

\noindent\textbf{Intent.} To allow synchronization between the two traces letting them evolve together which otherwise would not be possible because of behavioral mismatches.
\\
\noindent\textbf{(1) MESSAGE CONSUMER PATTERN.}
\\
\noindent\textbf{Problem.} (1) Extra send/missing receive mismatch ((1) in Figure \ref{FIG:basic_mismatches}, where the extra send action is $\overline{message_2}$). One of the two considered traces either contains an extra send action or a receive action is missing.
\\
\noindent\textbf{Example.} Consider two traces implementing the abstract action ``send (respectively receive) message''. For example, in the mismatch (1) of Figure \ref{FIG:basic_mismatches} the right trace implements only the sending of the message while the left trace implements the receiving of the message and the sending of an acknowledgment ($\overline{message_2}$).
\\
\noindent\textbf{Solution.} Introducing a \textit{message consumer} (solution of mismatch (1) in Figure \ref{FIG:basic_mediator_patterns}) that is made by an action that, ``consumes'' the identified extra send action by synchronizing with it, letting the two traces communicate.
\\
\noindent\textbf{Example Resolved.} First the two traces synchronize on the sending/receiving of the message ($message_1$) and then the left trace synchronizes its sending of the acknowledgment ($\overline{message_2}$) with the message consumer that receives it.
\\
\noindent\textbf{Variants.} Possible variants and respective solutions are represented in Figure \ref{FIG:var_basic_mismatch_1} and are:
\begin{itemize}
\item[(a)] $message_1$ has exchanged send/receive type within the two traces, i.e., the left trace is the sequence $\overline{message_1}.\overline{message_2}$ while the right trace is just $message_1$. In this case the message consumer remains the same ($message_2$).

\item[(b)] $\overline{message_1}$ is the extra send message instead of $\overline{message_2}$. The left trace is the sequence $\overline{message_1}.\overline{message_2}$ while the right trace is made by $message_2$. In this case the message consumer performs $message_1$.

\item[(c)] the extra send message is $\overline{message_1}$, the left trace is the sequence $\overline{message_1}$$.$ $message_2$ while the right trace is $\overline{message_2}$. In this case the message consumer is made by $message_1$.
\end{itemize}
%
\noindent\textbf{(2) MESSAGE PRODUCER PATTERN.}
\\
\noindent\textbf{Problem.} (2) Missing send/extra receive mismatch ((2) in Figure \ref{FIG:basic_mismatches}, where the missing send action is $\overline{message_2}$). One of the two considered traces either contains an extra receive action or a send action is missing in it. This is the dual problem of mismatch (1).
\\
\noindent\textbf{Example.} Consider two traces implementing the abstract action ``send (respectively receive) message''. In the mismatch (2) of Figure \ref{FIG:basic_mismatches}, the right trace implements the sending of the message ($\overline{message_1}$) and the receiving of an acknowledgment ($message_2$) while the left trace implements just the receiving the message ($message_1$).
\\
\noindent\textbf{Solution.} Introducing a \textit{message consumer} (solution of mismatch (2) in Figure \ref{FIG:basic_mediator_patterns}) made by an action that ``produces'' the  missing send action corresponding to the identified extra receive action and let the two traces synchronize.
\\
\noindent\textbf{Example Resolved.} The two traces first synchronize on the sending/receiving of the message ($message_1$) and then the right trace synchronize its receive of the acknowledgment ($message_2$) with the message consumer mediator that sends it.
\\
\noindent\textbf{Variants.} Possible variants and respective solutions are shown in Figure \ref{FIG:var_basic_mismatch_2} and are:
\begin{itemize}
\item[(a)] $message_1$ has exchanged send/receive type within the two traces, i.e., the left trace is $\overline{message_1}$ while the right trace is the sequence $message_1.message_2$ In this case the message producer performs $\overline{message_2}$.

\item[(b)] the missing send message is $\overline{message_1}$, instead of being $message_2$, the right trace is the sequence $message_1.\overline{message_2}$ while the left trace is made by $message_2$. In this case the message producer is made by $\overline{message_1}$.

\item[(c)] the missing send message is $message_1$, the left trace is $message_2$ while the right trace is the sequence $message_1$$.$$\overline{message_2}$. In this case the message producer is made by $\overline{message_1}$.
\\
\end{itemize}
%
\noindent\textbf{(3) MESSAGE TRANSLATOR PATTERN.}
\\
\noindent\textbf{Problem.} (3) Signature mismatch (upper right box of Figure \ref{FIG:basic_mismatches}). The two traces represent semantically complementary actions but with different signatures. With signature we mean only the action name.
\\
\noindent\textbf{Example.} Consider two traces implementing the abstract action ``send (respectively receive) information request''. Instantiating the mismatch (3) of Figure \ref{FIG:basic_mismatches}, $\overline{message_1}$ could be the send of a message {\em Information} while $message_2$ the receive of a {\em Request} message.
\\
\noindent\textbf{Solution.} Introducing a {\it message translator} (solution of mismatch (3) in Figure \ref{FIG:basic_mediator_patterns}). It receives the request and sends it after a proper translation. We assume the existence of some entity able to do the translation\footnote{Technically the message translator synchronizes twice with the involved components using different messages and this implements a translation.}. Referring to the example, the translator mediator trace is: $message_1$$.$$\overline{message_2}$.
\\
\noindent\textbf{Example Resolved.} First the message {\em Information} is exchanged between one trace and the mediator. After its translation, a {\em Request} message is sent by the mediator to the other trace. The message translator performs: {\em Information}.$\overline{Request}$.
\\
\noindent\textbf{Variants.} A possible variant with its solution is shown in Figure \ref{FIG:var_basic_mismatch_3} and amount to exchange sender/receiver roles between the two traces, i.e., $message_1$ and $\overline{message_2}$ and the solution is made by $message_2.\overline{message_1}$.
\\
%
\noindent\textbf{(4) MESSAGES ORDERING PATTERN.}
\\
\noindent\textbf{Problem.} (4) Ordering mismatch ((4) in Figure \ref{FIG:basic_mismatches}, where both traces perform complementary (send/receive) $message_1$ and $message_2$ but in different order). Both traces consist of complementary functionalities but they perform the actions in different orders. Nevertheless this mismatch can be considered also as a combination of extra/missing send/receive actions mismatches (1) and (2), however we choose to consider it as a first class mismatch. Generally speaking, it may happen that not all the ordering problems are solvable due to the infinite length of the traces. However this is not our case.
\\
\noindent\textbf{Example.} Consider two traces implementing the abstract action ``send (respectively receive) name''. $message_1$ and $message_2$ in the mismatch (4) of Figure \ref{FIG:basic_mismatches}, for example, correspond to $FirstName$ and $LastName$ respectively. Then, one sends the sequence $FirstName$$.$$LastName$ while the other receives $LastName$$.$$FirstName$.
\\
\noindent\textbf{Solution.} Introducing a {\it messages ordering} (solution of mismatch (4) in Figure \ref{FIG:basic_mediator_patterns}). This pattern has a compatible behavior for both the traces. The pattern is made by a trace that receives the messages and, after a proper reordering, resends them.
\\
\noindent\textbf{Example Resolved.} Referring to the example, the messages ordering trace is: $message_1.message_2$ $.$ $\overline{message_2}$ $.$ $\overline{message_1}$ that ~is ~$FirstName$$.$$LastName$$.$$\overline{Last}$- $\overline{Name}$$.$$\overline{FirstName}$.
That is, first one trace synchronizes with the mediator which receives the messages and then the mediator reorders the messages and sends them to the other trace.
\\
\noindent\textbf{Variants.} Possible variantsand respective solutions are shown in Figure \ref{FIG:var_basic_mismatch_4} and are:
\begin{itemize}
\item[(a)] left trace has exchanged sender/receiver role with respect to the right trace, i.e., the left trace is the sequence $\overline{message_2}$$.$$\overline{message_1}$ while the right trace is the sequence $message_1$$.$$message_2$. In this case the messages ordering is the sequence $message_2.message_1$$.$$\overline{message_1}.\overline{message_2}$.

\item[(b)] in both traces the first action is a send while the second is a receive. That is, the left trace is $\overline{message1}$$.$$message2$ while the right is $\overline{message2}$$.$$message1$. In this case the message ~ordering ~is ~the ~sequence $message_1$$.$$message_2$$.$ $\overline{message_2}$$.$$\overline{message_1}$.

\item[(c)] in both traces the first action is the receive followed by the send. That is, the left trace is $message_1$ $.$ $\overline{message_2}$ while the right is $message_2$$.$$\overline{message_1}$. In this case the basic solution to solve the mismatch is not the messages ordering. It is a proper combination of messages producers and consumers (message producer followed by message consumer for the left trace followed by message producer followed by message consumer for the right trace). That is, $\overline{message_1}$$.$$message_2$ followed by $\overline{message_2}$$.$$message_1$.
    \\
\end{itemize}
%
\noindent\textbf{(5) MESSAGE SPLITTING PATTERN.}
\\
\noindent\textbf{Problem.} (5) One send-many receive/many receive-one send mismatch ((5) in Figure \ref{FIG:basic_mismatches}).
The two considered traces represent a semantically complementary functionality but one expresses it with one action and the other with two actions.
\\
\noindent\textbf{Example.} Consider two traces implementing the abstract action ``send (respectively receive) name''. Instantiating the one send-many receive mismatch (5) of Figure \ref{FIG:basic_mismatches}, for example, $\overline{message}$ can be the send of one message $\overline{FirstLastName}$ while $message_1$ and $message_2$ are the receive of two separate messages $FirstName$ and $LastName$.
\\
\noindent\textbf{Solution.} Introducing a {\it messages splitting} (solution of mismatch (5) in Figure \ref{FIG:basic_mediator_patterns}). It receives one message from one side, splits it properly, and sends the split messages to the other. We assume the existence of some entity able to do the splitting operation\footnote{Technically the message splitting synchronizes several times with the involved components using different messages and this implements a split.}. Referring to the example, the trace of the message splitting (5) is: $message$$.$$\overline{message_1}$$.$$\overline{message_2}$.
\\
\noindent\textbf{Example Resolved.} With respect to the example, the mediator first performs one receive, then a splitting, and subsequently sends two messages. That is, of $FirstLastName$$.$$\overline{FirstName}$$.$$\overline{LastName}$.
\\
\noindent\textbf{(6) MESSAGE MERGER PATTERN.}
\\
\noindent\textbf{Problem.} (6) Many send-one receive/one receive-many send mismatch ((6) in Figure \ref{FIG:basic_mismatches}).
The two considered traces represent a semantically complementary functionality but they express it with a different number of actions. This is the dual problem of mismatch (5).
\\
\noindent\textbf{Example.} Consider two traces implementing the abstract action ``send (respectively receive) name''. Instantiating the many send-one receive (6) of Figure \ref{FIG:basic_mismatches}, for example, $\overline{message_1}$ and $\overline{message_2}$ are the sending of two separate messages $\overline{FirstName}$ and $\overline{LastName}$ while $message$ is the receiving of $FirstLastName$.
\\
\noindent\textbf{Solution.} Introducing a {\it messages merging} (solution of mismatch (6) in Figure \ref{FIG:basic_mediator_patterns}). It receives two messages from one side, merges them properly, and sends the merged messages to the other. We assume the existence of some entity able to do the merge operation. Referring to the example, the trace of the messages merging is: $message_1$$.$$message_2$$.$$\overline{message}$.
\\
\noindent\textbf{Example Resolved.} With respect to the example, the mediator first performs two receives, then a merge, and subsequently sends one message. That is, $FirstName$$.$$LastName$$.$$\overline{FirstLastName}$.
\section{Application of the Patterns to the example}
\label{SEC:application}

The aim of this section is to show the patterns at work, putting together all the jigsaws puzzle.
Thanks to the compatibility analyzer, we discover that the two messengers components are potentially compatible, i.e., our instant messengers share some intent having complementary portion of interaction protocols. Hence, it makes sense to use the architectural Mediating Connector Pattern to mediate their conversations.
Following the pattern-based approach described in Section \ref{SEC:approach}, the messengers behavior is decomposed into traces representing elementary behaviors. Then, the traces are analyzed and their basic mismatches are identified thanks to the basic mediators patterns.
Subsequently, a composition strategy is applied to build elementary mediators, i.e., mediator traces, exploiting the basic mediators patterns.
Finally, in order to automatically synthesize the behavior of the whole Mediating Connector for the messengers, a composition approach aggregates the elementary mediators so to have a mediated coordination and communication.

\begin{figure}[htbp]
\centering \epsfig{file=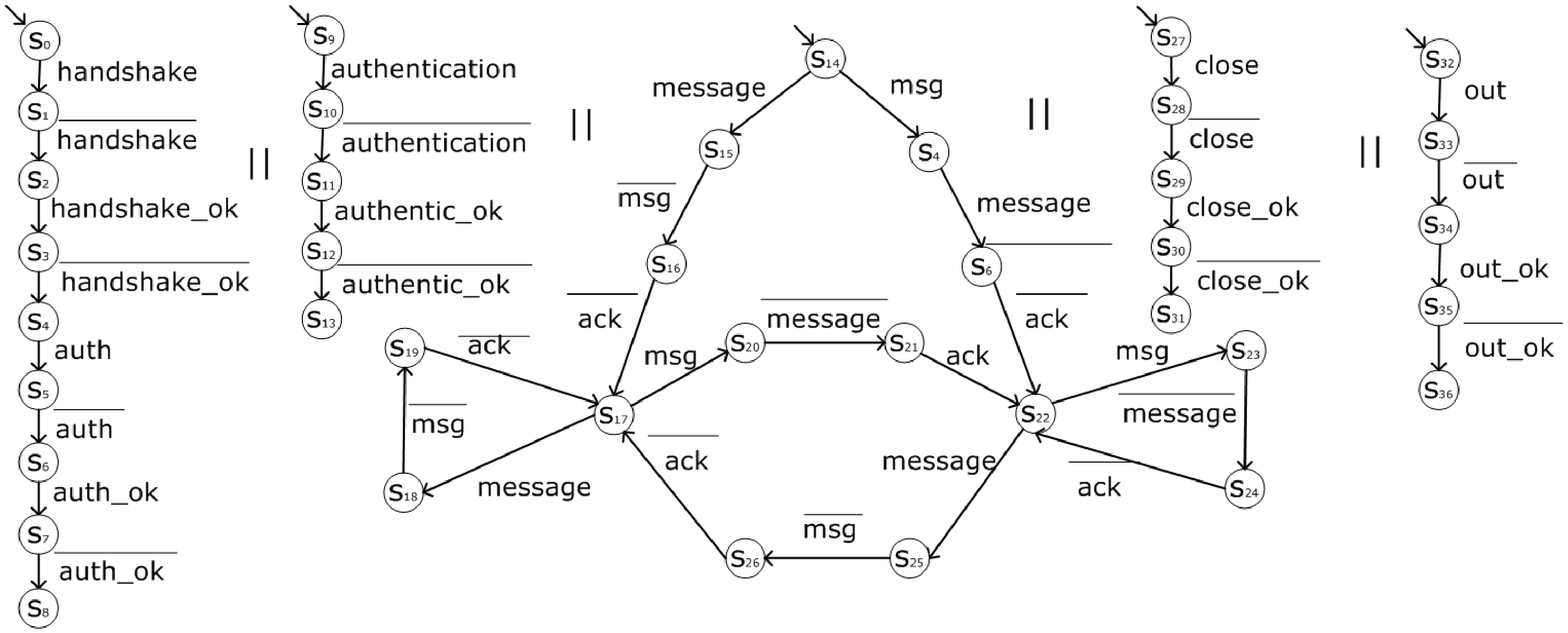,
width=4.9in}
 \caption{Behavioral description of the Mediating Connector for the messengers example}
\label{FIG:connector}
\end{figure}

Figure \ref{FIG:connector} shows the behavior of the Mediating Connector for the messengers.
%
The mediator, in this example, allows the interaction between the two different messengers by translating conversations from one protocol into the other and forwarding them while only forwarding the interactions between each messenger and its server.
In particular, the Windows Messenger uses acknowledged messages while the Jabber Messenger does not use acknowledges. This is detected and solved by the mediator thanks to the application of the Message Consumer Pattern and Message Producer Pattern, from which messages ``$\overline{ack}$'' and ``$ack$'' originate respectively.
\section{Related work}
\label{SEC:related}
In the last decades the notion of mediators has been investigated in several contexts. Initially introduced to cope with the integration of heterogeneous data sources \cite{WIEDERHOLD},\cite{WIEDERHOLD2} and as design pattern \cite{GOF}, it has been subsequently studied within the Web context to cope with many heterogeneity dimensions \cite{STOLLBERG} and with behavioral mismatches that may occur during interactions between business processes (protocol mediation).
An approach to protocol mediation is to categorize the types of protocol mismatches that may occur and that must be solved in order to provide corresponding solutions to these recurring problems.
This immediately reminds of patterns whose pioneer was Alexander \cite{ALEXANDER}.
Patterns have received attention in several research areas: for example Bushmann et al. \cite{BUSCHMANN} have defined patterns for software architectures; more recently in \cite{PARIS} a pattern language has been proposed revisiting existing architectural patterns; and the ``gang of four'' in \cite{GOF} have defined design patterns.
Between all, two design patterns are related to our: the Mediator Design Pattern, which is a behavioral pattern, and the Adapter which is structural. The first one is similar because it serves as intermediary for coordinating the interactions among groups of objects but it is different because its main aim is to decrease the complexity of interactions. The second is similar because it adapts the interfaces of the objects while it differs because our mediator is not just an interface translator.

In the Web services context several works have introduced basic pattern mismatches and the corresponding template solutions to help the developers to compose mediators \cite{BENATALLAH,CIMPIAN,CIN2,CIN3}.
The Web Services research community has been also studying solutions to the automatic mediation of business processes in recent years \cite{VACULIN07,VACULIN08,SEMI07,WILLIAMS}.
A lot of work has been also devoted to behavioral adaptation that is related to our work among which, for instance, \cite{BENATALLAH2}. It proposes a matching approach based on heuristic algorithms to match services for the adapter generation taking into account both the interfaces and the behavioral descriptions. Our matching, as sketched in Section \ref{SEC:approach}, is driven by the ontology and is better described in \cite{SII_WICSA_ECSA09} where the theory underlying our approach is described at a high level and in \cite{ISOLA_10_THEO} where a more detailed version of the theory can be found.
The work \cite{DUMAS} presents an algebra over services behavioral interfaces to solve six mismatches and a visual notation for interface mapping. The proposed algebra describes with a different formalism solutions similar to our basic patterns LTSs and this can be of inspiration for us in the direction of the reasoning.
Finally in \cite{TSE_CANAL} the authors propose an approach for software adaptation which takes as input components behavioral interfaces and adaptation contracts and automatically builds an adaptor such that the composed system is deadlock-free. The adaptor generation is also tool supported. Similarly to us, they also solve some mismatches.
A difference, instead, is that our aim is to achieve communication without adding extra constraints while they say that their main goal is to ensure deadlock freedom and addresses system-wide adaptation specified through policies and properties.

\section{Conclusion}
\label{SEC:conclusion}

The Ubiquitous environment, embedding a big number of heterogeneous system's components, puts forward an ever growing need of mediation entities for component's interoperability purpose.
The challenge is to embed {\it mediators} components able to solve the component's behavioral discrepancies, into the system architecture allowing components to coordinate and communicate.
Moreover, due to the continuous evolution of such environments, recently has also emerged the issue of dynamic and on the fly solutions to behavioral diversities letting hence arise the challenge of automatic approaches to cope whit this problem.

To respond to these two challenges, we first illustrated the Mediating Connector Architectural Pattern which, encapsulating the necessary support, is the key enabler for the communication between mismatching components. Indeed, it solves the interoperability problems between heterogeneous and potentially compatible components.
With respect to the second challenge, we proposed an automatic pattern based approach detailing, in particular, a set of Basic Mediator Patterns, including basic mismatches and respective solutions, which represent the basic building blocks on which an automatic approach can build upon.

As future works, we intend to: define a theoretical compositional strategy to allow reasoning on mismatches and to build the mediating connector behavior. Moreover we also aim at providing the ``concrete'' Basic Mediator Patterns, i.e., the skeleton code corresponding to the ``abstract'' ones presented in this work and present the actual code for the component's behavior decomposition and the mediating connector behavior building.

%

\bibliographystyle{eptcs} 

\end{document}